\documentclass[9pt,twocolumn,twoside]{pnas-new}
\templatetype{pnasresearcharticle} %

\title{Thermal conductivity of B-DNA}

\author[a,1]{Vignesh Mahalingam}
\author[a,2]{Dineshkumar Harursampath}

\affil[a]{Department of Aerospace Engineering, Indian Institute of Science, Bangalore, 560012, India}

\leadauthor{Mahalingam}

\significancestatement{The thermal conductivity of B-form double-stranded DNA (dsDNA) of the Drew-Dickerson sequence d(CGCGAATTCGCG) is computed using classical Molecular Dynamics (MD) simulations. In contrast to previous studies, which focus on a simplified 1D model or a coarse-grained model of DNA to improve simulation times, full atomistic simulations are employed to understand the thermal conduction in B-DNA. Thermal conductivity at different temperatures from 100 to 400 K are investigated using the Einstein Green-Kubo equilibrium and Müller-Plathe non-equilibrium formalisms.}

\authorcontributions{Vignesh Mahalingam performed research and Vignesh Mahalingam, Dineshkumar Harursampath wrote the paper together}
\authordeclaration{The authors declare no conflict of interest.}
\correspondingauthor{\textsuperscript{2}To whom correspondence should be addressed. E-mail: vigneshm\@iisc.ac.in}


\begin{abstract}
The thermal conductivity of B-form double-stranded DNA (dsDNA) of the Drew-Dickerson sequence d(CGCGAATTCGCG) is computed using classical Molecular Dynamics (MD) simulations. In contrast to previous studies, which focus on a simplified 1D model or a coarse-grained model of DNA to improve simulation times, full atomistic simulations are employed to understand the thermal conduction in B-DNA. Thermal conductivity at different temperatures from 100 to 400 K are investigated using the Einstein Green-Kubo equilibrium and Müller-Plathe non-equilibrium formalisms. The thermal conductivity of B-DNA at room temperature is found to be 1.5 W/m$\cdot$K in equilibrium and 1.225 W/m$\cdot$K in non-equilibrium approach. In addition, the denaturation regime of B-DNA is obtained from the variation of thermal conductivity with temperature. It is in agreement with previous works using Peyrard-Bishop Dauxois (PBD) model at a temperature of around 350 K. The quantum heat capacity ($\mathbf{C_{vq}}$) has given the additional clues regarding the Debye and denaturation temperature of 12-bp B-DNA.
\end{abstract}

\dates{This manuscript was compiled on \today}
\doi{\url{www.pnas.org/cgi/doi/10.1073/pnas.XXXXXXXXXX}}

\begin{document}

\maketitle
\thispagestyle{firststyle}
\ifthenelse{\boolean{shortarticle}}{\ifthenelse{\boolean{singlecolumn}}{\abscontentformatted}{\abscontent}}{}

\dropcap{P}robing thermal conduction in shorter length scales using computations gives us a fundamental understanding of the link between structure and phenomena. DNA satisfies the low thermal conductivity requirements for building molecular thermoelectric devices~\cite{nianprb2019}, This was a motivation for this study. In an experiment to find the thermal conduction of DNA-gold composite where $\mathrm{\lambda}$-DNA coated with gold nanoparticles~\cite{Kodama2009}, the thermal conductivity ($\kappa$) was found to be 150 W/m$\cdot$K. The later works have cited that gold primarily contribute to this high value of thermal conductivity and have mentioned that the ultra-low thermal conductivity of DNA~\cite{Xuaip2014,Xupolymer2014,SavinPRB2011}. An experiment using a new transient electro-thermal technique with crystalline DNA composite fibers in NaCl solution, has suggested that this might indeed be the case as the thermal conductivity found for the fibers are low around 0.25-0.85 W/m$\cdot$K~\cite{Xuaip2014,Xupolymer2014}. 

The earlier computational works use the Peyrad-Bishop-Dauxois (PBD) model which is a simplified 1-D non-linear bead spring model~\cite{Peyrardiop2004}. This model has been used to understand DNA dynamics and to find the denaturation point of DNA~\cite{PeyrardPRL1989}. The denaturation regime of DNA was found to be around 350 K and thermal conductivity of the DNA in ~\cite{VelizhaninPRE2011} at 300 K is 1.8 mW/m$\cdot$K. This value is much lower than that of poly(G) DNA ($\approx$ 0.3 W/m$\cdot$K) obtained using a 12-coarse grained (12-CG) model~\cite{SavinPRB2011}. The disagreement raises the question of the validity of thermal conductivity computed using PBD model. A few infinite chains formed by permuting Adenine (A) and Guanine (G) sequences-poly(A), poly(G), poly(AG), poly($\mathrm{A_2G_2}$), poly($\mathrm{A_{200}G_{200}}$) are investigated using PBD model and it was found that the denaturation point can be shifted depending on the sequence~\cite{Chieniop2013}. Additionally, the thermal conductance ratio $\mathrm{R=\kappa_H/\kappa_L}$, which is defined as the high thermal conductivity to low thermal conductivity of a various sequences is quantified to analyze thermal switching in DNA. The sequence poly($\mathrm{A_2G_2}$) seemed to have higher the thermal conductance ratio than poly(A) and poly(G) sequences. Again, the thermal conductance values were extremely lower, probably owing to the PBD model~\cite{Chieniop2013}. The PBD model has also been used to understand the improving heat conduction through external force~\cite{Behniaepjb2016}. As an extension of this, DNA switching was studied by considering one end of DNA helical turn as drain, another end as source and the central region as gate\cite{Behniacp2018}. An all-atom picture resolves the discrepancies in understanding the fundamental nature of thermal conduction in a DNA. Hence, the water and $\mathrm{Na^+}$ ions are considered as the stabilizing media and only the thermal conductivity of B-DNA is calculated.

\begin{figure}%
  \includegraphics[width=\linewidth]{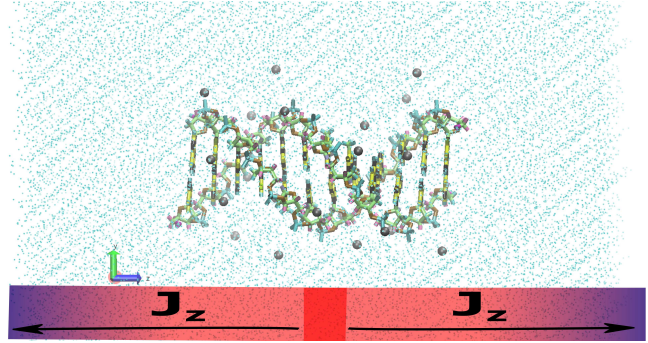}
  \caption{The simulated B-DNA structure is in a water box with $\mathrm{Na^+}$ ions. The B-DNA has a total length of 40.8 \AA \ and a radius of 10 \AA \ and is colored according to different atom types. Also shown is the non-equilibrium M\"{u}ller-Plathe scheme where the red region with higher temperature swaps kinetic energies with blue regions with lower temperatures.}
  \label{fgr:fig1}
\end{figure}
\section*{Results and discussion}
\subsection*{Thermal conductivity using Green-Kubo equilibrium method}
\begin{figure}%
  \includegraphics[width=\linewidth]{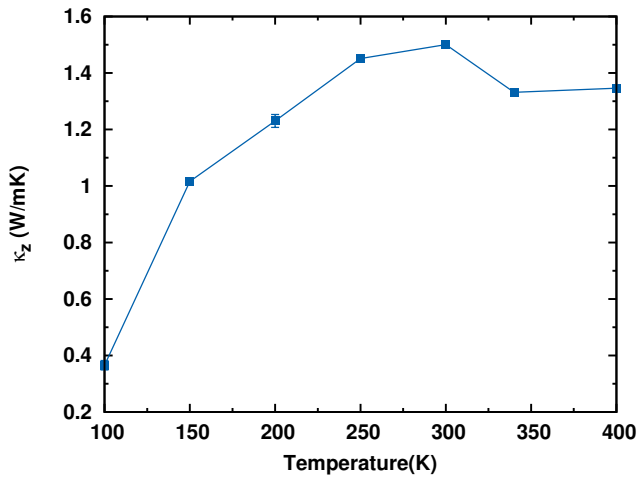}
  \caption{Thermal conductivity of B-DNA along the length of DNA as a function of temperature~\cite{Xupolymer2014}.}
  \label{fgr:fig2}
\end{figure}

The thermal conductivity of B-DNA is calculated using the equilibrium GK method as in equation (\ref{eqn:eqn1}). This requires the calculation of heat-heat auto-correlation function (HCAF) from heat current using equation (\ref{eqn:eqn2}) and this settles to an equilibrium value. The thermal conductivity value obtained at 300 K along the length of DNA after it has settled to an equilibrium value for 10 ns. All equilibrium GK thermal conductivity values at different temperature are obtained similarly. The thermal conductivity of the dsDNA sequence along its length as a function of temperature is shown in Figure~\ref{fgr:fig2}. The thermal conductivity increases as a function of temperature and eventually saturates around 350 K. 

\begin{figure}%
  \includegraphics[width=\linewidth]{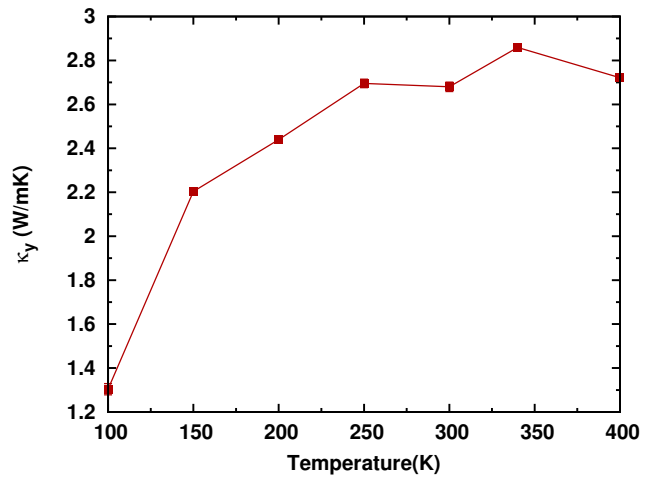}
  \caption{Thermal conductivity of B-DNA along base-pairs $\kappa_y$ as a function of temperature.}
  \label{fgr:fig6}
\end{figure}

The equilibrium method also allows us to calculate the heat flux along the other directions such as those between the strands as shown in Figure~\ref{fgr:fig6}. These thermal conductivities are inaccessible to experiments at such shorter strand lengths. Interestingly, the heat transfer along base pairs between backbone is higher than that along the strand. No parallels exist in reported literature about the heat conduction between strands, although ~\cite{SavinPRB2011} has mentioned that the heat conduction along the length is primarily due to sugar-phosphate backbone. The classical treatment here has not accounted for the transfer of heat between base pairs by tunneling. Nevertheless, it is evident from Figure~\ref{fgr:fig6} that more heat can be transferred along base pairs than along the length of phosphate backbone.

\subsection*{Power density spectrum}
\begin{figure}%
  \includegraphics[width=\linewidth]{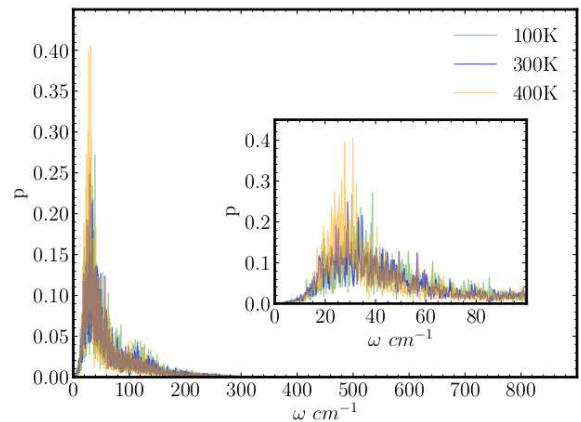}
  \caption{Power spectrum density of B-DNA at 300 K, 100 K and 400 K.}
  \label{fgr:fig3}
\end{figure}

To understand the molecular origin of the temperature dependence of the thermal conductivity of the DNA, the Density of States (DoS) of the 12-bp B-DNA  have been calculated using a 2-point (2pt) code~\cite{lin2003,lin2010,pascal2011} and have been plotted in Figure~\ref{fgr:fig3}. Only continuous low frequency modes can be seen till 800 cm$^{-1}$ (Debye frequency, $\omega_D$). In an earlier work on poly-G DNA~\cite{SavinPRB2011}, the DoS spectrum had a gap with no modes between 200 cm$^{-1}$ and 300 cm$^{-1}$. Moreover, few modes existed beyond this gap till 400 cm$^{-1}$.  No such gap is seen in the spectrum between optical and acoustic modes. No high frequency modes has been observed in both this work and in earlier work~\cite{SavinPRB2011}, meaning that the phonon modes have large wavelengths and hence are scattered at the dsDNA boundaries. Around the denaturation temperature, the DNA strands separate, and the thermal conductivity saturates. It is possible to compute the Debye temperature,$\mathrm{T_D}$ from this spectrum by using $\mathrm{T_D=\hbar\omega_D/k_B}$ , where $\hbar$ and $k_b$ are reduced Planck's constant and Boltzmann constant, respectively. Substituting $\omega_D$ to be 723 cm$^{-1}$ as it the last available frequency mode, $\mathrm{T_D}\ \approx$ 165 K.  

\begin{figure}%
\includegraphics[width=\linewidth]{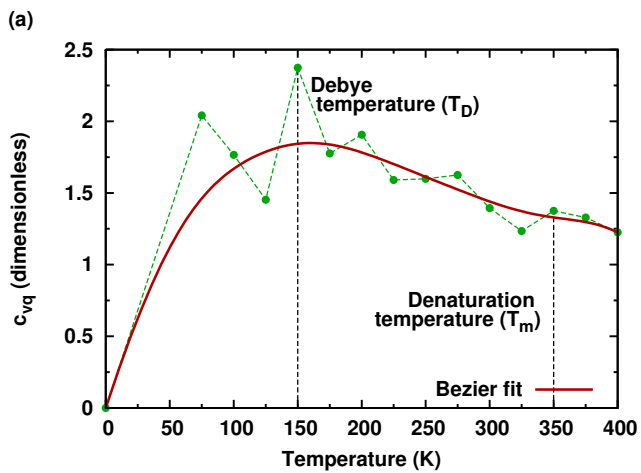}
\caption{Quantum heat capacity ($\mathrm{C_{vq}}$) of B-DNA as a function of temperature.}
\label{fgr:fig4}
\end{figure}

\begin{figure}
\includegraphics[width=\linewidth]{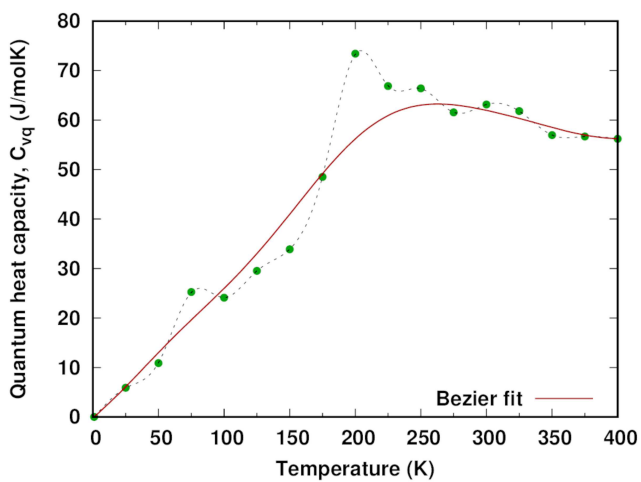}
\caption{Quantum heat capacity ($\mathrm{C_{vq}}$) of TIP3P water as a function of temperature.}
\label{fgr:fig5}
\end{figure}

Using the DoS shown in Figure~\ref{fgr:fig3}, also computed is the quantum molar specific heat capacity ($C_{vq}$) of the dsDNA (Figure~\ref{fgr:fig4}). The resulting heat capacity is plotted as a function of temperature. The peaks in heat capacity at 200 K and 273 K in Figure~\ref{fgr:fig4} are probably due to water as similar features can be seen in water heat capacity at the same temperatures in Figure~\ref{fgr:fig5}. The peak at 150 K is close to the calculated Debye temperature (165 K) of DNA. The peak around 350 K ought to correspond with the denaturation regime, where the transition from double strand to two single strands happens~\cite{wildesPRE2011}. It is till this point that the thermal conductivity increases and beyond which thermal conductivity saturates. A similar study~\cite{VelizhaninPRE2011} has described the same phenomenon and it is mentioned that the increase in thermal conductivity is strongly correlated with the anharmonicity in the bond between the complementary base pairs till there is effectively no contact between the complementary base pairs.

\subsection*{Thermal conductivity using Reverse Non-Equilibrium Molecular Dynamics (RNEMD) method}

\begin{figure}
\includegraphics[width=0.8\linewidth]{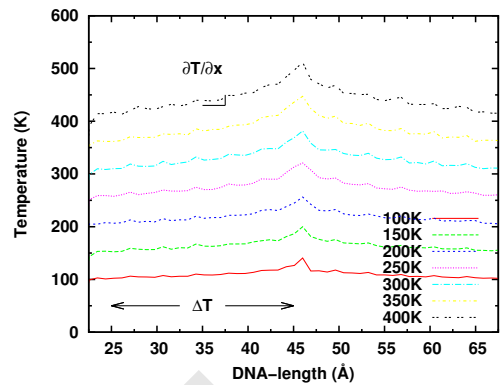}
\caption{Temperature profile across the water box which has B-DNA along z-direction}
\label{fgr:fig7}
\end{figure}

\begin{figure}
\includegraphics[width=0.8\linewidth]{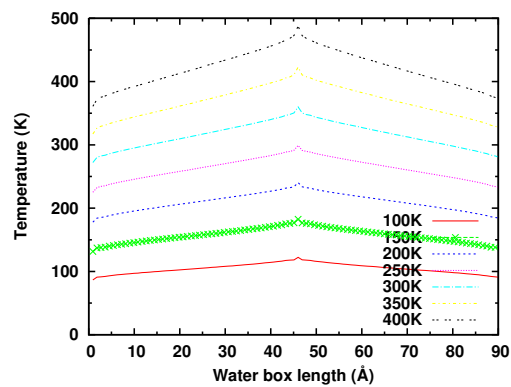}
\caption{Temperature profile across the 12-bp B-DNA in RNEMD simulation along z-direction}
\label{fgr:fig8}
\end{figure}

A non-equilibrium MP approach~\cite{mp1997} is also used to understand the low thermal conductivity obtained earlier using equilibrium formulation. Here, a temperature gradient can be set along the length of B-DNA and surrounding water box. The temperature profile across the surrounding water clearly has a gradient between the center hot region and cold regions on either side. Figures~\ref{fgr:fig7} and \ref{fgr:fig8}  shows the water and DNA temperature profiles, respectively.  

\begin{figure}%
  \includegraphics[width=\linewidth]{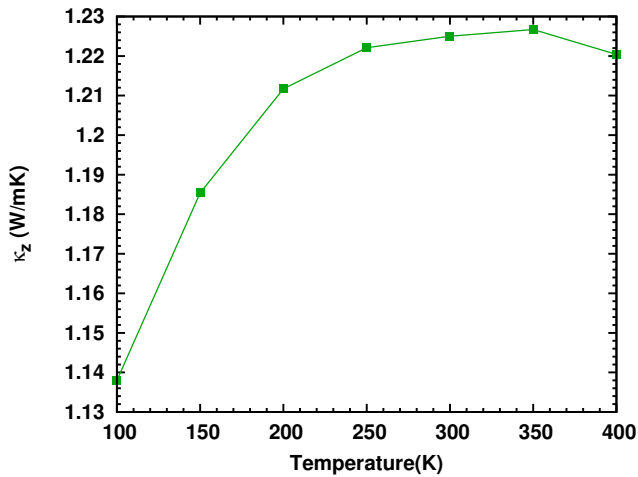}
  \caption{Thermal conductivity versus temperature for 12-bp B-DNA in RNEMD simulation. Inset shows the Green-Kubo values plotted in Figure ~\ref{fgr:fig2}.}
  \label{fgr:fig9}
\end{figure}

The thermal conductivity $\kappa$ due to a linear temperature gradient between the DNA ends (between 22.5 and 67.5 \AA ) is given as
 \begin{equation}
\kappa_{z}= \frac{\left( \frac{Q}{At} \right)}{\left(  \partial T / \partial z \right)},
\label{eqn:eqn9}
\end{equation}
where Q is the heat exchange between hot and cold regions, A is the cross-sectional area of the water box and t is the time for heat exchange. The temperature gradient is computed across B-DNA from the temperature profile in Figure ~\ref{fgr:fig8}. Caution was exercised in calculating the temperature profile as constrained SHAKE atoms were excluded and a mild Berendsen thermostat was used~\cite{zhang2005}. Figure~\ref{fgr:fig9} shows the temperature dependence of the thermal conductivity using this method and the profile similar to GK thermal conductivity (Figure ~\ref{fgr:fig2}). 

Till denaturation there is an increase in thermal conductivity $\mathrm{\kappa}$ and heat capacity (C) of B-DNA with respect to temperature. This is attributed to increase of phonon density with respect to temperature \cite{Xupolymer2014} as  $\mathrm{\kappa=\frac{Cv^2\tau_r}{3}}$  in the single time relaxation approximation. Moreover, it can be seen from Figure~\ref{fgr:fig3} that only low frequency (long wavelength) soft modes are available in DNA allowing this classical approximation to be valid. The heat capacity (C) of B-DNA also increases as Debye law states: $C_{vq} \propto T^3$ (Refer Figure~\ref{fgr:fig4}). It is being assumed that the phonon velocity (v) remains almost constant with temperature and the relaxation time $\tau_r\propto T^{-1}$~\cite{Xupolymer2014}. Hence, the thermal conductivity initially increases almost quadratically with respect to temperature till the DNA strands separate. The thermal conductivity of the DNA-gold composite is found to be 150 W/m$\cdot$K ~\cite{Kodama2009}, whereas recent estimates of the thermal conductivity of DNA-water composite mixture via equilibrium and non-equilibrium MD were 0.381 W/m$\cdot$K and 0.373 W/m$\cdot$K, respectively~\cite{nitasha2020}. Our results suggest that there is a definite contribution of gold and water to the thermal conductivity of the composite mixture in these works and the thermal conductivity of DNA is somewhere close to the reported values in this work.

\section{Discussion}
We have examined the thermal conductivity of 12-bp B-DNA, the most common form of DNA from GK calculation  from RNEMD calculation. Both calculations show an increase in thermal conductivity till denaturation temperature. The full atom description as opposed to coarse grained or 1D non-linear chain models indicate the regions where the models succeed and fail. The thermal conductivity obtained using PBD models needs to be refined as they seem to be very low. Nevertheless, all models predict the denaturation regime close to 350 K, where the thermal conductivity saturates with increase in temperature. The 2pt-calculations show that the Debye temperature is consistent with the earlier works~\cite{VelizhaninPRE2011,SavinPRB2011,Chieniop2013}. The engineering of thermal conductivity, based on the base-pair sequences along the long lengths, might play a role in its usage as a molecular thermoelectric device operating at room temperature. This work lays the foundation for an all-atom study of DNA thermal conductivity. Building further using the methods here would give us insight into the dependence of thermal conductivity on base-pair sequences from all-atom perspective. The thermal conductivity computed here might be a necessary validity check for coarse-grained DNA models. 

\matmethods{All the calculations are performed on the 12-base pair (bp) B-DNA of Drew-Dickerson sequence d(CGCGAATTCGCG)~\cite{Dickerson475}.  Nucleic Acid Builder (NAB) module of AMBERTOOLS18~\cite{caseamber18} is used to build the initial structures of the double stranded (ds) DNA. The dsDNA is then placed in a bath of TIP3P water box~\cite{Jorgensenjcp1983} using xleap module of AMBERTOOLS18 software package. A water box with dimensions of 65 \AA \ $\times$ 65 \AA \ $\times$ 90 \AA \ (\textit{x}$\times$\textit{y}$\times$\textit{z}) is chosen to ensure 15 \AA \ solvation shell around the B-DNA. 22 $\mathrm{Na^+}$ ions are added at the lowest electrostatic potential locations to the solvated dsDNA system. DNA OL15 force-field~\cite{Zgarbovajctc2015} is used. This has parmbsc0~\cite{PEREZ2007} and OL15~\cite{Zgarbovajctc2015} corrections to the ff99 force-field~\cite{Wangjcc2000} to consider the bonded and non-bonded interactions of the dsDNA. Joung-Cheathem parameter~\cite{Joungjpc2008} set is used to consider the interaction of monovalent $\mathrm{Na^+}$ ions with TIP3P water and dsDNA. 
After preparing the initial system using AMBERTOOLS18~\cite{caseamber18}, LAMMPS~\cite{PLIMPTON19951} software module is used for all further simulations. The whole solvated dsDNA system is energy minimized using first 5000 steps of steepest descent and 5000 steps of conjugate gradient keeping the B-DNA restrained with a force of 500 kcal/mol\AA . The DNA is then slowly released into water by reducing the restraint from 20 kcal/mol\AA \ to 0 kcal/mol\AA \ in 5 cycles of the steepest descent and conjugate gradient minimization steps. All the atoms are then assigned velocities according to Maxwell-Boltzmann distribution. Throughout the MD simulation, the DNA has a small restraint of 1 kcal/molÅ to prevent the same from changing its orientation whilst measuring thermal conductivity. SHAKE constraints~\cite{RYCKAERT1977} are applied to the hydrogen atoms, bond and angles of DNA and water with a tolerance of $\mathrm{10^{-4}}$\ ~\cite{Wangjcc2000}. The system is equilibrated for 10 ns with Nos\'{e}-Hoover thermostat and barostats with coupling constants 0.1 ps and 1.0 ps, respectively~\cite{shinoda2004rapid,martyna1994constant,parrinello1981polymorphic,tuckerman2006liouville}.  Finally, a production run of 20 ns for the calculation of thermal conductivity ensures that the thermal conductivity values converge. The thermal conductivity is computed using the equilibrium Green-Kubo (GK) method, where the heat-heat auto-correlation function is used to compute the thermal conductivity as~\cite{green1952,green1954,kubo1957}: 

\begin{equation}
\kappa_{x,y,z}= \frac{V}{k_BT^2} \int_0^\infty \left< J_{x,y,z} (0) \cdot J_{x,y,z} (t) \right> dt,
\label{eqn:eqn1}
\end{equation}

where thermal conductivity $\kappa_{x,y,z}$ at a temperature T in a direction x, y or z is obtained from heat current $\mathrm{J}_{x,y,z}$ in that direction. $k_B$ is the Boltzmann constant. The heat current is obtained as 

\begin{equation}
\mathbf{J}= \frac{1}{V} \left[ \sum_i e_iv_i - \sum_i \sigma_iv_i \right]
\label{eqn:eqn2}
\end{equation}

where  $e_i$ is the total energy of an atom, $v_i$ is the velocity of an atom, $\sigma_i$ is the virial stress per atom and V is the volume of the total group of atoms. The exact volume (V) of B-DNA is computed from the atomic volumes of adenine (136.1 \AA$^{3}$), guanine ((143.8 \AA$^{3}$), cystosine (113.2 \AA$^{3}$)and thymine (132.6 \AA$^{3}$) base-pair groups and sugar-phosphate (174.8 \AA$^{3}$) groups~\cite{Nadassy2001,voss2005}. Each strand (left or right of B-DNA symmetrical axis) contains 4 cytosine, 4 guanine, 2 adenine and 2 thymine and 12 sugar-phosphate groups giving us the total volume of the 12-bp B-DNA to be 7326 \AA$^{3}$. The power spectrum density or Density of States (DoS) of the 12-bp B-DNA is obtained from a fast Fourier transform of velocity-velocity auto-correlation, C(t) as

\begin{equation}
DoS(\nu)= \lim\limits_{t \rightarrow \infty} \frac{1}{2k_BT} \int\limits_{-\tau}^\tau C(t) e^{-2\pi \nu t} dt,
\label{eqn:eqn3}
\end{equation}
 
where t is correlation time window of 200 ps and $\nu$ is the frequency. Only the solid component of DoS is considered as liquid and gaseous states are not relevant for B-DNA. The canonical partion function (Q) can be constructed from DoS, with a harmonic oscillator assumption:

\begin{equation}
\ln Q = \int\limits_0^\infty DoS(\nu) q_{HO}(\nu) d\nu,
\label{eqn:eqn4}
\end{equation}

where $q_{HO}= \frac{e^{-\beta h\nu}}{1-e^{-\beta h\nu}}$ is the harmonic oscillator partition function, $\beta=\frac{1}{k_BT}$ and $h$ is the Planck's constant. The entropy $S^0$ and the heat capacity $C_{vq}$ are then found using the partition function and DoS as
\begin{equation}
S^0= k \ln Q + \beta^{-1} \left(\frac{\partial \ln Q }{\partial T}\right)_{N,V} = k \int\limits_0^\infty DoS(\nu) W^s (\nu) d\nu
\label{eqn:eqn5}
\end{equation}
\begin{align}
C_{vq} = \left( \frac{\partial S^0}{\partial T} \right)  
= k \left(\frac{\partial \ln Q }{\partial T}\right)_{N,V} + \beta^{-1} \left(\frac{\partial^2 \ln Q }{\partial T^2}\right)_{N,V} \nonumber \\
= k^2\beta^{-2} \int\limits_0^\infty DoS(\nu) W^C (\nu) d\nu \qquad \qquad \qquad \qquad \qquad
\label{eqn:eqn6}
\end{align}
with weighting functions

\begin{align}
W^s (\nu) = \frac{\beta h \nu}{e^{\beta h\nu}-1} \ \ln\left[1-\left(e^{-\beta h\nu}\right)\right], \\
W^C (\nu) = \frac{e^{-\beta h\nu}}{\left[ 1-\left( e^{-\beta h\nu}\right) \right] ^2}. \qquad \qquad \quad
\end{align}

}

\showmatmethods{} %

\acknow{The authors thank Dr. Navaneetha Krishnan, Dr. Prabal Maiti and Abhishek Aggarwal for their fruitful discussions and suggestions.}

\showacknow{} %

\bibliography{pnas-main}

\end{document}